\documentclass[11pt,letterpaper]{article}
\usepackage[utf8]{inputenc}
\usepackage[T1]{fontenc}
\usepackage{lmodern}
\usepackage{microtype}
\usepackage[margin=1in]{geometry}
\usepackage{etoolbox}
\usepackage{amsmath}
\usepackage{graphicx}
\usepackage{xcolor}
\usepackage{booktabs}
\usepackage{array}
\usepackage{tabularx}
\usepackage{float}
\usepackage{changepage}
\usepackage{caption}
\usepackage[numbers,sort&compress]{natbib}
\usepackage[hidelinks]{hyperref}
\usepackage{xurl}
\newcolumntype{L}{>{\raggedright\arraybackslash}X}
\newlength{\extralength}
\newlength{\fulllength}
\makeatletter\g@addto@macro\@floatboxreset\centering\makeatother

\graphicspath{{figures/}}
\newcommand{\ascp}[1]{\textrm{Ascon-}p[#1]}
\AtEndPreamble{\pdfstringdefDisableCommands{\def\,{}}}

\newcommand{\paperTitle}{Replay-Protected Authenticated Encryption
for CAN~FD with Ascon-AEAD128 and Flash-Persisted Freshness on FPGA}

\newcommand{\authorA}{Kaviprashanna Krishnamoorthy}
\newcommand{\affilA}{Independent Researcher}
\newcommand{\orcidIdA}{0009-0002-0478-0422}
\newcommand{\authorB}{Vyshnav Nandha}
\newcommand{\orcidIdB}{0009-0006-0705-0504}
\newcommand{\affilB}{Texas A\&M University, College Station, TX, USA}
\newcommand{\emailB}{nandhavyshnav@gmail.com}

\newcommand{\paperAbstract}{CAN~FD frames are transmitted in cleartext, so any node on the bus can read, forge or replay them. We present a hardware secure layer for CAN~FD that encrypts and authenticates each frame with NIST Ascon-AEAD128 and rejects replays using a 48-bit freshness counter that persists across reboots. Each 64-byte data field carries a 46-byte payload, the low 16 bits of the counter and a full 128-bit tag. The receiver reconstructs the full counter from these 16 bits and uses it to form the nonce, so tag verification also rejects replayed frames. Each endpoint reserves counter values in blocks and commits the upper bound of each block to SPI NOR flash before use. In simulation, power interruptions injected during flash writes and erases never caused nonce reuse. On an Artix-7 XC7A35T at 100~MHz, the layer occupies 2549 lookup tables (LUTs) alongside the publicly available CTU CAN~FD controller. Its simulated processing overhead is about 1.3~$\mu$s per frame, under 1\% of the frame's airtime. We evaluated it on a physical CAN~FD bus with both nodes on one FPGA. An independent adapter recorded highway torque commands and replayed them during braking. The unprotected receiver acted on every replay. The secure layer rejected every replay and still accepted every genuine command.}

\newcommand{\paperKeywords}{CAN FD; Ascon; AEAD; NIST SP 800-232; replay attack;
freshness; SecOC; FPGA; automotive security}

\hypersetup{pdftitle={\paperTitle},
  pdfauthor={\authorA\ and \authorB},
  pdfsubject={cs.CR; cs.AR},
  pdfkeywords={\paperKeywords}}
\title{\paperTitle}
\author{%
  \authorA$^{1}$ \qquad \authorB$^{2,}$\thanks{Corresponding author:
  \texttt{\emailB}}\\[8pt]
  \parbox{0.9\textwidth}{\centering\small
  $^{1}$\,\affilA, Bengaluru, India\\
  ORCID \href{https://orcid.org/\orcidIdA}{\orcidIdA}\\[4pt]
  $^{2}$\,\affilB\\
  ORCID \href{https://orcid.org/\orcidIdB}{\orcidIdB}}}
\date{\normalsize Preprint, November 2026}

\begin{document}
\maketitle
\begin{abstract}
\noindent\paperAbstract
\end{abstract}
\medskip
\noindent\textbf{Keywords:} \paperKeywords

\section{Introduction}\label{sec:intro}

A modern vehicle relies on dozens of electronic control units (ECUs), and much of the traffic between them still runs on the Controller Area Network (CAN)~\cite{iso11898}. CAN assumes that every node on the bus is trusted. Its frames are sent in cleartext and carry no proof of the sender's identity. A receiver also has no way to tell a new frame from a replayed one.

An attacker who gains electrical access to a bus segment, for example through a compromised ECU or the diagnostic port, can read all traffic and inject frames, including replays of recorded ones~\cite{woo2016tits,lotto2024survey}. Message authentication alone does not prevent replay, because a recorded frame still carries a valid authenticator and can be resent without knowledge of the key. A secure layer must therefore also check freshness and reject any frame it has already accepted. This record of accepted frames must persist across reboots, since an attacker able to interrupt power could otherwise reset it and replay old traffic.

We implement such a layer in hardware on a field-programmable gate array (FPGA), using a publicly available CAN~FD controller and a standardized cipher. CAN~FD extends the data field from 8 to 64 bytes, enough for a 46-byte payload, a two-byte counter and a full 128-bit tag. The 18 bytes of security fields alone would not fit in the 8-byte data field of classic CAN. For the cipher we use Ascon-AEAD128, the lightweight mode for authenticated encryption with associated data (AEAD) that NIST standardized in SP~800-232 in August 2025~\cite{nist2025sp800232}. The practice of transmitting only part of the freshness counter comes from AUTOSAR Secure Onboard Communication (SecOC)~\cite{autosar2024secoc}. We adopt this practice, but we also encrypt the payload and keep the full 128-bit tag. We do not claim compliance with any SecOC profile. The layer operates at the controller's register interface and requires no processor. Encrypting or verifying a frame takes a fixed 61 clock cycles.

We use the cipher exactly as NIST specifies it. What is new is how it is put to work on a CAN~FD bus. Each part has precedent (Section~\ref{sec:related}), but we found no earlier FPGA design that combines these four:

\begin{enumerate}
\item \textbf{An integrated FPGA secure layer for CAN~FD.} We combine an Ascon-AEAD128 core with the publicly available CTU CAN~FD controller (release v2.7) on an Artix-7 XC7A35T FPGA running at 100~MHz. The secure layer uses 2549 lookup tables (LUTs). With one controller, it uses 4286 LUTs, not counting the logic that feeds frames to the controller. In simulation, the layer adds about 1.3~$\mu$s to each frame on average, sender and receiver combined.
\item \textbf{A derived nonce with authenticated freshness.} AEAD needs a nonce, a 128-bit value that must not repeat under one key. Both ends build it from a fixed device prefix and a 48-bit freshness counter. Each frame carries only the low 16 bits of that counter. The receiver keeps $L$, the last counter it accepted, and recovers the full value as the smallest counter above $L$ that matches those 16 bits. Since the tag is computed under the nonce, a valid tag also shows that the receiver recovered the counter the sender used. A replayed frame carries a counter at or below $L$, so the receiver forms the wrong nonce and the tag check fails. A frame that arrives after a newer one has been accepted is rejected in the same way. Up to 65\,535 consecutive frames can be lost without breaking synchronization. A formal proof by $k$-induction shows that the receiver accepts a counter only if it exceeds $L$ by at least 1 and at most 65\,536. The proof covers all counter values below a small guard band at the top of the 48-bit range.
\item \textbf{Reserve-ahead counter persistence.} Before an endpoint uses a counter value, it has already written a bound above that value to external NOR flash on a Serial Peripheral Interface (SPI) bus, with a new bound about every 1024 frames. After a reboot, each endpoint resumes above every counter value it used or accepted before, which prevents both nonce reuse and the replay of earlier frames. Every flash record carries a Berger check code. Under the fault model of NOR flash, in which an interrupted write or erase can only turn a record's zero bits into ones, this code detects any resulting corruption (Section~\ref{sec:persistence}). A receiver reboot costs at most 1536 genuine frames. Power loss during flash operations was tested in simulation only.
\item \textbf{A replay test on a physical bus with an independent adapter.} A separate CANable adapter, holding no key, recorded genuine frames carrying scripted torque commands and later replayed them unchanged. The unprotected receiver executed each replayed command. With the secure layer, the receiver dropped each replayed frame and still executed all genuine commands.
\end{enumerate}

The sender and the receiver hold the same 128-bit key, which is installed in advance and never exchanged or renewed over the bus. In the prototype, each endpoint has its own CAN~FD controller and transceiver, but both run on one FPGA and share the rest of the board, including its clock, flash device and power supply. All experiments were carried out on a lab bench.

\section{Background and Related Work}\label{sec:background}

\subsection{CAN and CAN~FD}\label{sec:canfd}

Every node on a CAN bus sees every frame. Each frame opens with an arbitration field holding the identifier, which names the message. When two nodes transmit at once, the frame with the lower identifier goes through. Nothing stops a node from using another node's identifier, and the cyclic redundancy check (CRC) uses no key. So a receiver cannot tell who sent a frame or whether it was sent before.

Classic CAN carries at most 8 data bytes and runs the whole frame at one bit rate, at most 1~Mbit/s. CAN with Flexible Data-Rate (CAN~FD), standardized in ISO 11898-1:2015~\cite{iso11898}, allows 64 data bytes, signaled by the data length code (DLC) 0xF. Its FD format indicator (FDF) bit marks the frame as CAN~FD, and its Bit-Rate Switch (BRS) bit lets the data phase run faster than arbitration. Classic CAN networks that use security mostly use AUTOSAR SecOC~\cite{autosar2024secoc}. Its security fields have to fit into the same 8 bytes as the payload (Section~\ref{sec:secoc}).

\subsection{Ascon-AEAD128 (NIST SP~800-232)}\label{sec:ascon}

Ascon-AEAD128~\cite{nist2025sp800232} uses a 128-bit key $K$ and a 128-bit nonce $N$, and appends a 128-bit tag $T$ to the ciphertext. Its 320-bit state consists of five 64-bit words, $S = S_0 \| S_1 \| S_2 \| S_3 \| S_4$, and is updated by $\ascp{n}$, a permutation of $n$ rounds. The data is XORed into the first 128 bits of the state, called the rate, and the ciphertext is read from there. Initialization and finalization run $\ascp{12}$, and $\ascp{8}$ separates consecutive data blocks. The standard also limits each key: nonces must not repeat, decryption may fail at most $2^{96}$ times, and at most $2^{54}$ bytes may be processed, nonces included (Section~\ref{sec:keylimits}).

\subsection{AUTOSAR SecOC}\label{sec:secoc}

AUTOSAR SecOC~\cite{autosar2024secoc} authenticates messages between ECUs but does not encrypt them. Each message carries a message authentication code (MAC) and the low bits of a freshness value, from which the receiver rebuilds the full value. On classic CAN both must fit beside the payload in 8 bytes, so the MAC is cut short: Profiles~1 and~3 send only 24 and 28 bits of an AES-based MAC, while the specification itself cites 64 bits as sufficient. How the freshness value survives a reboot is left to the implementation.

\subsection{Related Work}\label{sec:related}

Nguyen and Chen~\cite{nguyen2024ascon} built Ascon v1.2 accelerators for classic CAN on an FPGA. Their Ascon-128 uses a 64-bit rate and is not the standardized cipher. In their bus demonstration the CAN data travels in plaintext, and no replay defense is described. Gladis Kurian and Chen~\cite{gladiskurian2025ascon} apply Ascon-AEAD128 from the initial public draft of SP~800-232 to Internet of Things (IoT) links and detect replays with a Bloom filter of nonces already seen. In their simulation of 100\,000 nonces, the filter wrongly rejects 0.166\% of genuine ones. It is also lost at power-off, and every message must carry the full nonce.

Several Ascon cores for FPGAs already exist. SystemVerilog implementations of the final standard are publicly available~\cite{primas2025asconverilog,lasconip2026}, and Alharbi et al.~\cite{alharbi2024ascon7series} evaluate a unified v1.2 core on the 7-series family used here. Choi and Moon~\cite{choi2026access} use final-standard Ascon-AEAD128 with a counter-derived nonce in an Ethernet gateway. Their receiver does not check for replays and keeps no record of the counters it has seen.

Earlier FPGA work on CAN~FD builds encryption and integrity from separate primitives. Doan and Ganesan~\cite{doan2017saecrypto} combine AES-128 with keyed SHA-1 and leave the replay counter to future work. Lee et al.~\cite{lee2023ikeee} combine AES-128 with HMAC-SHA-256 in an engine of 11\,761 LUTs on an Artix-7. Their two ends keep a 256-bit counter in step, but the paper gives no way to resynchronize it and does not store it. Lagnf and Ganesan~\cite{lagnf2021eit} encrypt an 8-bit counter with the payload, but their design was only simulated and has no CAN controller. Athira et al.~\cite{athira2022ijraset} only add an unkeyed SHA-256 digest, which any node on the bus can recompute. Gao~\cite{gao2024ecie} only encrypts.

Rosenstatter et al.~\cite{rosenstatter2019freshness} describe how SecOC Profile~3 keeps a trip counter in non-volatile memory, and they propose a Profile~4 that resynchronizes on demand. In our design the FPGA logic itself writes the counter bounds to external flash, with no processor involved, and a receiver reboot costs at most 1536 genuine frames.

Lotto et al.~\cite{lotto2024survey} survey software authentication for classic CAN, where the MAC must be truncated or sent in extra frames to fit the 8-byte data field. CINNAMON~\cite{bella2020cinnamon} extends the AUTOSAR SecOC module with encryption. Rasheed et al.~\cite{rasheed2024tnse} encrypt with TinyJAMBU-128 and detect replays from message context, with detection rates of 0.92 to 0.99. For CAN~FD, Woo et al.~\cite{woo2016tits} propose a software security architecture, and Watada et al.~\cite{watada2026icct} run the pre-standard Ascon-128 on Raspberry Pi nodes. Instead of Ascon's AEAD mode, they build their own encryption mode and MAC from the Ascon permutation. The 28-bit sequence number wraps to one on overflow. Key renewal by elliptic-curve Diffie-Hellman (ECDH) lets them rekey in the field. Our keys can be changed only when the ECU is serviced, although a key-epoch byte in the associated data leaves room for field rotation (Section~\ref{sec:limits}).

For CAN~XL, the next CAN generation, CAN in Automation (CiA) has released CANsec as a draft specification, CiA 613-2~\cite{cia613-2}. It secures the data link layer in the style of Ethernet's MACsec. Wiemer and Zeh~\cite{wiemer2023cansec} analyze its cipher suites, recommend adding Ascon and note that ``for Classic CAN and CAN~FD a native security protocol was not considered.'' For CAN~FD itself, the SPsec/CANcrypt V2 project~\cite{spsec2026} defines a security sublayer with 64-bit tags, but its documentation states that it is not specified for use in motor vehicles.

\section{Design}\label{sec:design}

\subsection{Threat Model}

The adversary controls the bus. It can read, modify, delay, drop, inject and replay frames, and it can power-cycle any node whenever it chooses. It does not know the pre-shared key and cannot break the cipher. It cannot take over a node or read the key stored in it. A node whose key has been extracted counts as compromised, and so does one whose flash has been rewritten or rolled back. The stored counter bounds carry only an unkeyed check code, so a forged record would be accepted.

The design does not protect availability. An adversary can stop traffic by flooding the bus or resetting a node, and repeated transmitter resets can leave the channel desynchronized (Section~\ref{sec:limits}). Side-channel attacks are out of scope.

\subsection{Architecture and Frame Format}\label{sec:arch}\label{sec:frame}

\begin{figure}[htbp]
\includegraphics[width=13.5cm]{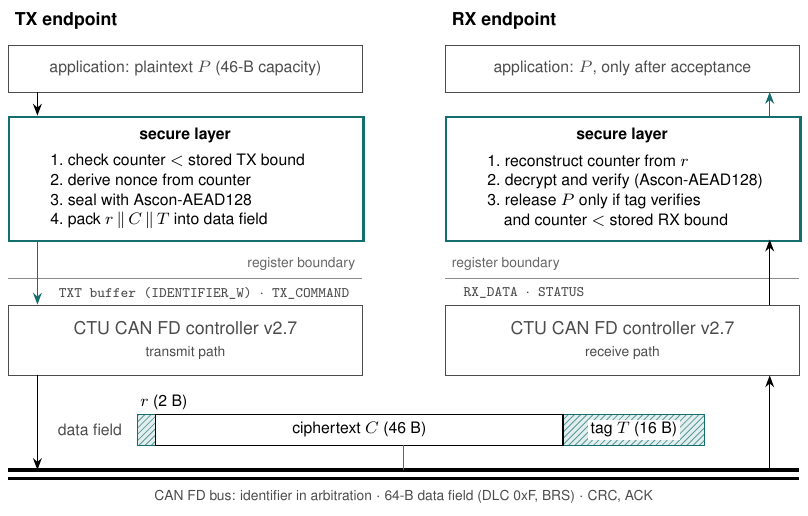}
\caption{System architecture: the secure layer sits
between the application and the CTU CAN~FD controller's register
interface at both ends. The data field is the low 16 counter
bits $r$, the ciphertext $C$ and the tag $T$, concatenated ($\|$). The evaluated build places both endpoints on one FPGA with a shared
clock, reset, test harness and flash.\label{fig:system}}
\end{figure}

The transmitter encrypts each 46-byte payload under a nonce built from its counter (Figure~\ref{fig:system}, Section~\ref{sec:freshness}). The 64-byte data field carries the ciphertext with its tag and the low 16 counter bits. The receiver rebuilds the full counter from those bits and accepts the payload only if the tag verifies and the counter is below the receiver's stored bound (Section~\ref{sec:persistence}). Table~\ref{tab:params} lists the design parameters.

\begin{table}[htbp]
\caption{Design parameters of the evaluated build.\label{tab:params}}
\begin{tabularx}{\textwidth}{lX}
\toprule
\textbf{Parameter} & \textbf{Value} \\
\midrule
Cipher & Ascon-AEAD128, SP~800-232 final; 128-bit tag, no truncation \\
Key & 128-bit public test key (deployment must provision a
secret); key epoch \texttt{0x00} \\
Nonce & derived: $\mathrm{DevicePrefix}(80) \cdot 2^{48} +
\mathrm{counter}(48)$, serialized little-endian \\
Counter & 48-bit monotonic; low 16 bits sent in the frame \\
Frame & fixed 64-byte data field (46-byte payload), DLC = 0xF, FDF+BRS \\
Associated data & 5 bytes = identifier word (4 bytes) $\|$ key epoch (1 byte) \\
Freshness & counter advance of 1 to $2^{16}$ per accepted frame;
assumes last accepted counter $L < 2^{48}-2^{16}$ (operating
condition, no runtime guard) \\
Persistence & per-direction reserve-ahead bound,
STRIDE 1024, in SPI NOR flash (Pmod SF3) \\
\bottomrule
\end{tabularx}
\end{table}

Every secure frame uses all 64 bytes. Shorter payloads are padded to 46 bytes before encryption. Messages of varying length would carry their own length inside the payload. The 18 bytes of security fields leave 71.9\% of the frame for payload. Sending the full nonce, as Gladis Kurian and Chen~\cite{gladiskurian2025ascon} do, would take 16 bytes instead of two and cut that share to 50\%. Halving the tag would free eight more bytes but raise the chance of a forgery from $2^{-128}$ to $2^{-64}$ per attempt, so the design keeps the full tag.

The tag is also computed over the frame's CAN identifier, so a frame copied to another identifier fails verification. Ascon takes this input as associated data: the controller's 4-byte identifier word in big-endian order, followed by a 1-byte key epoch reserved for key rotation (Section~\ref{sec:limits}). Neither is sent in the data field, since the identifier is already in the arbitration field and the epoch is configured locally.

\subsection{Derived Nonce and Authenticated
Freshness}\label{sec:freshness}

\begin{figure}[htbp]
\includegraphics[width=13.5cm]{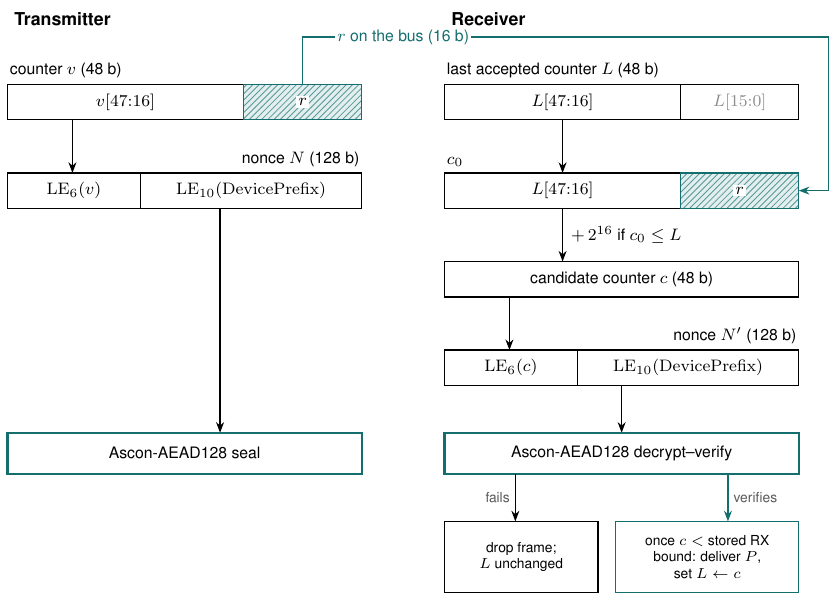}
\caption{Nonce derivation at the transmitter and single-candidate
counter reconstruction at the receiver
(Equation~\ref{eq:reconstruct}). The striped teal cells
are $r$, the 16 counter bits carried on the bus. In the little-endian byte order of SP~800-232, $N = \mathrm{LE}_6(v) \| \mathrm{LE}_{10}(\mathrm{DevicePrefix})$, each value written least significant byte first.\label{fig:nonce}}
\end{figure}

Each transmitter numbers its frames with a 48-bit counter $v$. The nonce is this number with the transmitter's fixed 80-bit prefix in front, $N = \mathrm{DevicePrefix} \cdot 2^{48} + v$, so it changes with every frame. Only the last 16 bits of $v$, called $r$, are sent (Figure~\ref{fig:nonce}), and the receiver works out the rest. MACsec~\cite{ieee8021aebw}, DTLS~1.3~\cite{rfc9147} and LoRaWAN~\cite{lorawan11} also send only the end of their counters. RFC~5116~\cite{rfc5116} allows this, because a wrongly recovered nonce makes the tag check fail.

The receiver remembers $L$, the counter of the last frame it accepted. After a reboot it starts from its stored bound in flash. Because the counter only moves forward, the receiver can fill in the missing upper bits from $L$. It first assumes the new counter shares the upper 32 bits of $L$. If that gives a value not above $L$, the low 16 bits must have rolled over, so it adds $2^{16}$:
\begin{equation}
c_0 = L[47{:}16] \,\|\, r, \qquad
c =
\begin{cases}
c_0 + 2^{16} & \text{if } c_0 \le L,\\
c_0 & \text{otherwise.}
\end{cases}
\label{eq:reconstruct}
\end{equation}
For example, if $L$ ends in 0xFFF0 and $r$ is 0x0005, the low bits have wrapped past 0xFFFF, so $c$ gets the next value of the upper bits of $L$. In every case, $c$ is the first counter after $L$ that ends in $r$. The receiver checks the tag under the nonce built from $c$ and drops the frame if it fails, leaving $L$ unchanged. Otherwise it delivers the frame once $c$ is below its stored bound (Section~\ref{sec:persistence}), and $L$ moves up to $c$.

The receiver thus recovers any counter from $L + 1$ to $L + 2^{16}$, so up to 65\,535 frames can be lost in a row, including values a reboot skips (Section~\ref{sec:persistence}). Replays are rejected. An old frame has a counter at or below $L$, so the receiver reconstructs the wrong counter and the tag does not verify. The same happens to a genuine frame that arrives more than $2^{16}$ counter values ahead, and the prototype cannot resynchronize after that. A forged tag passes with probability $2^{-128}$ per attempt (Section~\ref{sec:keylimits}). The window enforces order only: a delayed frame that was never accepted still goes through when it arrives.

The counter cannot run to the very end of its 48-bit range. Close to $2^{48}$, adding $2^{16}$ overflows, and an old frame would verify again under its original nonce. Equation~(\ref{eq:reconstruct}) is therefore valid only for $L < 2^{48} - 2^{16}$, and the proof uses the slightly lower bound $2^{48} - 2^{17}$ as a guard band. The prototype does not check for this. In practice the counter never gets there: SP~800-232 caps the data per key, and that cap forces a new key sooner (Section~\ref{sec:keylimits}).

\subsection{Counter Persistence}\label{sec:persistence}\label{sec:persistimpl}

If a reboot reset the counter, the transmitter would reuse nonces under the same key, which can expose the plaintext~\cite{nist2025sp800232}. So the counter has to survive reboots, for confidentiality as much as for replay protection. We follow the approach in RFC~5116~\cite{rfc5116}: write a value some steps ahead to non-volatile memory, then use the values below it.

Both the transmitter and the receiver store a bound $h$ in flash, a ceiling the counter must stay below. A new, higher $h$ is always written before the counter gets close. After a reboot, each side starts again at $h$. The transmitter has never encrypted with a value that high, and the receiver has never accepted one, so nothing from before the reboot can repeat. The receiver needs its own bound because only it knows which counters it accepted.

The bound advances in blocks of 1024 counter values ($\mathrm{STRIDE} = 1024$), so flash is written only about once every 1024 frames. The next bound is written when the counter is halfway through the current block, which leaves time for the write to finish. If it has not finished when the counter reaches the bound, the counter pauses, and the waiting frame is sent or accepted once the write is done. The receiver raises its counter only for frames whose tag verifies, so forged frames cannot push the bound forward or wear out the flash by forcing writes.

A receiver reboot discards some genuine frames. On restart, the receiver sets $L$ to its stored bound, which may lie up to $\mathrm{STRIDE} + \mathrm{STRIDE}/2 = 1536$ above the last counter it accepted, and frames with counters in that range are rejected. This bound assumes that the counter advances by one per frame and that each new bound reaches flash in time. A smaller STRIDE of 16 lowers the bound to 24 frames but requires more frequent flash writes.

The bounds are stored in an external NOR flash connected over SPI. Each direction has two 4~KB subsectors, A and B, each with room for 512 eight-byte records. A record holds a type byte, the 48-bit counter and a check byte. Records are written to A until it is full. Writing then continues in B, which is already erased, and A is erased in the background. At boot, the FPGA reads both subsectors and uses the highest counter with a valid check byte as $h$, so a record half-written during a power cut is ignored. A logic reset does not stop a flash operation already in progress, so the FPGA first waits for the flash to become idle, with no timeout.

The check byte holds the number of zero bits in the other seven bytes, a scheme known as a Berger code. If power fails while NOR flash is being programmed or erased, some bits that should read 0 may read 1, but no bit can go from 1 to 0. Damaged data therefore contains fewer zeros than the check byte says, and a damaged check byte can only show a larger count, so a damaged record never passes. Switching subsectors is also safe, because the newest valid record is never in the subsector being erased. Errors that also turn ones into zeros, such as random bit flips, could cancel out in the count and are not reliably detected.

Flash wear limits the lifetime. All four subsectors lie in one 64~KB sector of the flash, which is rated for 100\,000 erase cycles~\cite{micron2016n25q256a}. Assuming each subsector erase uses up one of those cycles, a channel lasts about $2.6 \times 10^{10}$ frames, or 8.3 years at 100~frames/s. The real figure is lower, because reboots and faults cause extra erases and the flash also has a data-retention limit. Moving the subsectors into separate sectors would roughly quadruple the lifetime. A larger STRIDE would extend it further, but each receiver reboot would then lose more frames.

A subsector erase can take up to 0.8~s, and while it runs the flash accepts no writes from either direction. Each new bound is requested with 512 counter values to spare, which is enough to cover an erase at up to about 640~frames/s. Above that rate, a direction whose bound is held up by an erase may pause briefly.

\subsection{Hardware Implementation}\label{sec:impl}\label{sec:core}\label{sec:platform}

\begin{figure}[htbp]
\includegraphics[width=13.5cm]{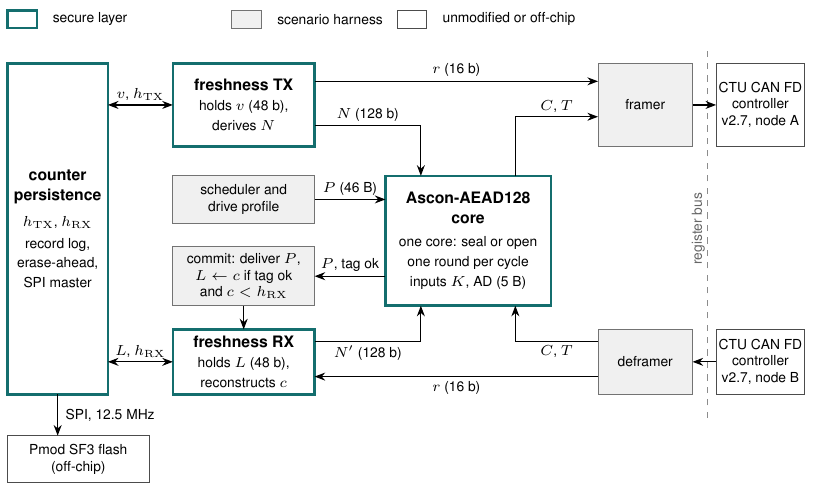}
\caption{Implementation datapath: transmit path along the top,
receive path along the bottom.\label{fig:rtl}}
\end{figure}

The secure layer is built from four blocks (Figure~\ref{fig:rtl}). An Ascon-AEAD128 core encrypts and tags outgoing frames, and verifies and decrypts incoming ones. A freshness module on each side manages the counter: the transmit side keeps $v$ and forms the nonce, and the receive side keeps $L$ and recovers each counter with Equation~(\ref{eq:reconstruct}). A persistence module stores both bounds in flash and reads them back at boot. Because the core takes the nonce as an input, the cipher can be tested apart from the freshness logic. A scenario harness drives the tests and schedules frames. In the prototype it also packs and unpacks the data field (Section~\ref{sec:frame}) and writes the controller registers.

\begin{figure}[htbp]
\includegraphics[width=0.8\textwidth]{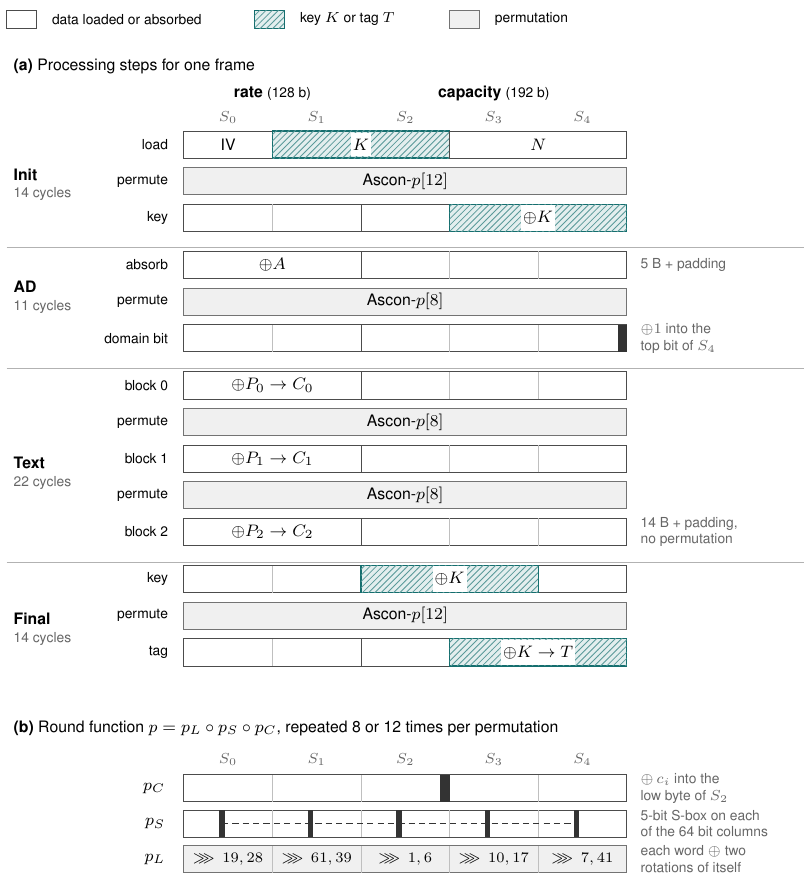}
\caption{Ascon-AEAD128 in the core for one frame. (a)~Each row
shows the state words that one step changes. The figure shows encryption; each ciphertext block stays in the
state, so the tag authenticates every block. (b)~The round function $p$ that each
permutation row repeats.\label{fig:fsm}}
\end{figure}

The core is written in SystemVerilog and computes one round of the Ascon permutation per clock cycle. Figure~\ref{fig:fsm}a shows how a frame passes through it. Initialization takes 14 cycles, the associated data 11, the 46-byte text 22 and finalization 14, for 61 cycles per frame. Four of those cycles are spent waiting for the harness to supply the next data block, so the core itself needs 57. The round function (Figure~\ref{fig:fsm}b) is ported from the \texttt{ascon-c} reference implementation~\cite{ascon2025refimpl}.

Decryption runs the same steps, but the plaintext leaves the core before the tag has been checked. The core reports the result only after finalization, with a flag that says whether the recomputed tag matches the one in the frame. In the prototype the harness holds the plaintext until this check and the flash-bound check both pass (Section~\ref{sec:arch}).

The secure layer talks to the CTU CAN~FD controller v2.7~\cite{ctucanfd} only through its registers, so any controller with the same register interface could take its place. We use the controller's source unmodified, apart from its build parameters. Its license permits research and evaluation use.

\begin{figure}[htbp]
\includegraphics[width=13.5cm]{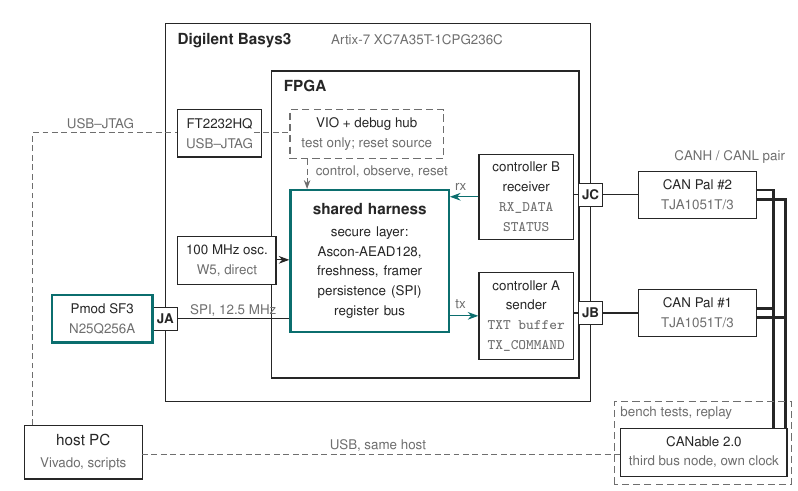}
\caption{Evaluation platform. Both endpoints and the shared
test harness run on one Basys3. The CANable is on the bus only during the
replay test.\label{fig:platform}}
\end{figure}

\begin{figure}[htbp]
\includegraphics[width=\textwidth]{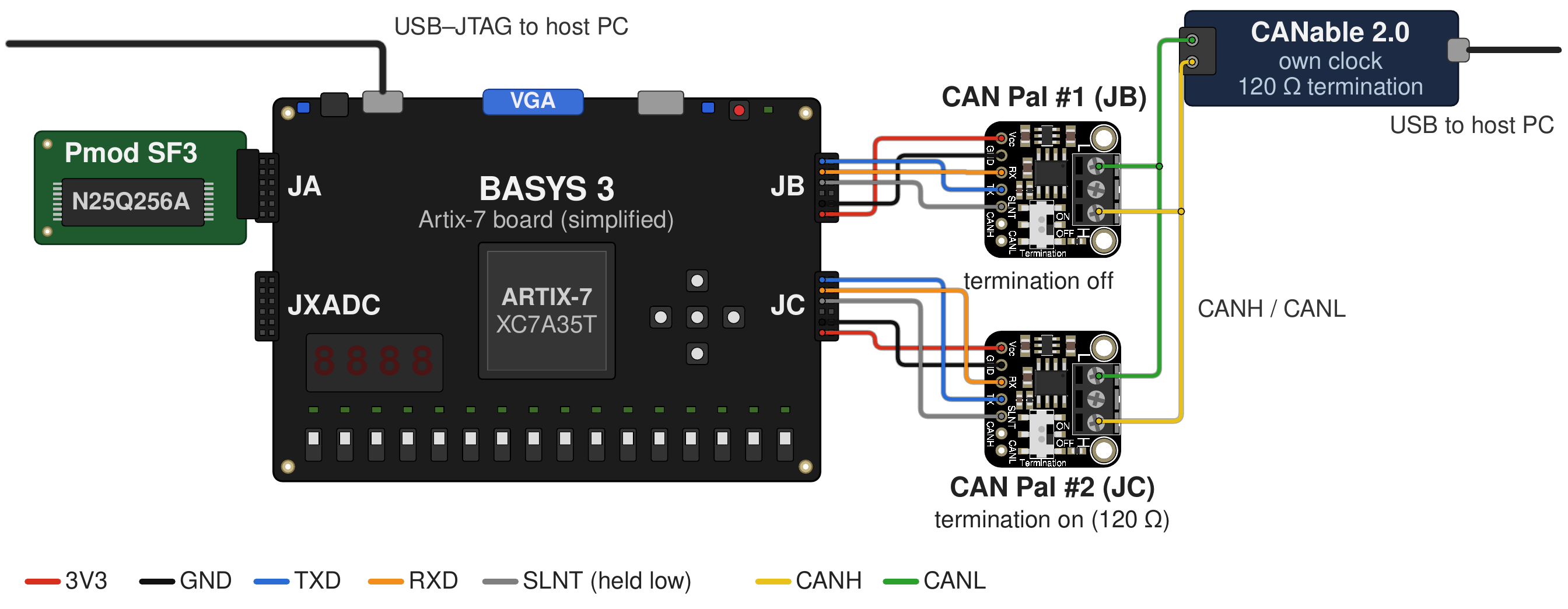}
\caption{Bench wiring (illustration, not a photograph). Each CAN
Pal is jumper-wired to its Pmod port: pin~1 TXD, 2 RXD, 3 SLNT
(held low by the FPGA), 5 GND, 6 3V3. CANH/CANL are taken from the
CAN Pals' screw terminals. The JC CAN Pal and the CANable
terminate the pair, and the JB CAN Pal is unterminated
(author-reported; resistance not measured). CAN Pal artwork: Adafruit Industries, Fritzing part
library, CC~BY-SA~3.0.\label{fig:bench}}
\end{figure}

The prototype runs on a Digilent Basys3 board with an Artix-7 XC7A35T and a 100~MHz oscillator (Figure~\ref{fig:platform}; wiring in Figure~\ref{fig:bench}). Each of the two controllers drives its own Adafruit CAN Pal transceiver (NXP TJA1051T/3), and a single CANH/CANL wire pair connects the two. The counter flash is a Digilent Pmod SF3 module (Micron N25Q256A), clocked at 12.5~MHz. Arbitration runs at 500~kbit/s and the data phase at 2 or 5~Mbit/s. The CANable adapter used in the replay test handles only 2~Mbit/s reliably, so 5~Mbit/s is used only between the two FPGA nodes (Section~\ref{sec:busresults}). On the receiving side, the controller's buffer is enlarged to 128 words to absorb replayed frames (Section~\ref{sec:limits}).

\section{Evaluation}\label{sec:verif}

\subsection{Verification in Simulation}\label{sec:bitexact}\label{sec:persistverif}

In simulation, the core gives the same output as the \texttt{ascon-c} reference (commit 446347f)~\cite{ascon2025refimpl} on all 1089 known-answer tests for SP~800-232. To test the tag check, we took 64 frames laid out as in our design and changed one bit at a time in the ciphertext, associated data, nonce or tag. All 42\,496 changed frames were rejected. A $k$-induction proof in SymbiYosys shows that the receiver accepts a counter only if it is 1 to $2^{16}$ above $L$ and ends in $r$, and that the nonce is built from it. The proof assumes $L$ is below the guard band (Section~\ref{sec:freshness}).

A Verilator regression runs the scenario harness of Section~\ref{sec:impl} against a bus model of the two controllers, reboots included. It checks that every counter used for encryption is below the transmitter's bound and never repeats. After each test in which the receiver rejects frames, it also checks that the receiver's bound is at most $\mathrm{STRIDE} + \mathrm{STRIDE}/2$ above $L$. Neither check fails in any run.

Power loss was tested in simulation, with a flash model that follows NOR behavior and can cut power at any point (Table~\ref{tab:persist}). No test produced a reused counter or an accepted replay. We also damaged 16\,000 records on purpose, and the Berger check rejected every one. A simple XOR checksum let 136 of 4000 such records through. Each of 15 deliberately faulty RTL versions failed at least one test.

\begin{table}[htbp]
\caption{Power-loss tests in simulation.\label{tab:persist}}
\begin{tabularx}{\textwidth}{>{\raggedright\arraybackslash}p{3.2cm}>{\raggedright\arraybackslash}X>{\raggedright\arraybackslash}X}
\toprule
\textbf{Test} & \textbf{Setup} & \textbf{Result} \\
\midrule
Reboot & 16 cuts to both ends together at $\mathrm{STRIDE} = 16$, 13 during an erase & all 1695 replays rejected; no genuine frame lost \\
Random cuts & 26 seeds of 3000 cuts during flash writes and erases & no counter reused; no receive bound moved back \\
\bottomrule
\end{tabularx}
\end{table}

\subsection{Replay on a Physical Bus}\label{sec:results}\label{sec:harness}\label{sec:boundeddemo}

The test models a replay attack on engine torque commands. Node A sends node B one command every 10~ms in the format of the J1939 Torque/Speed Control 1 message, following a fixed drive schedule. Highway acceleration requests 75\% torque and city braking requests 0\%, so a highway command replayed during braking asks for torque the driver did not request. A plaintext baseline sends the same commands without protection and acts on every one it receives.

The replays are sent by a CANable USB adapter, connected as a third node with its own clock. It records ten frames during highway acceleration and retransmits them unchanged during city braking. The adapter has no key.

In the plaintext baseline, all ten replayed frames reached the application during city braking, each requesting 75\% torque (Table~\ref{tab:replay}). With the secure layer, all ten failed verification and were dropped, and every genuine command was accepted. Neither run missed a slot or had a transport event (a transmit timeout or receive-buffer overrun).

\begin{table}[htbp]
\caption{Replay test on the physical bus at a 2~Mbit/s data rate.
Ten highway frames (75\% torque) are replayed unchanged during city
braking (0\%). The baseline's accepted count includes the ten
replays.\label{tab:replay}}
\begin{tabularx}{\textwidth}{lrrr>{\raggedleft\arraybackslash}X}
\toprule
\textbf{Mode} & \textbf{Genuine sent} & \textbf{Accepted} &
\textbf{Rejected} & \textbf{Positive torque in city phase} \\
\midrule
Plaintext baseline & 32\,978 & 32\,988 & 0 & 10 \\
Secure layer & 32\,978 & 32\,978 & 10 & 0 \\
\bottomrule
\end{tabularx}
\end{table}

To confirm that the hardware implements standard Ascon-AEAD128, we decrypted every captured secure frame offline with the \texttt{ascon-c} reference implementation. Every genuine frame decrypted to its scheduled command, and the counters had no gaps.

\subsection{Reboot and Power Loss}\label{sec:persistresults}

In a one-hour run on the evaluated build plus debug logic, both endpoints were reset together 289 times, about half of them during traffic, with the flash kept powered. All 7\,129\,630 frames were accepted. After every reset, the transmitter resumed above its last used counter and the receive bound never moved back. The 16-bit frame counter wrapped 113 times, and each direction switched subsectors about 14 times. A second run reset the endpoints 1100 times, 100 frames apart, to exercise every record position in the log. All 110\,000 frames were accepted.

We then switched the board off and on once, so that the flash also lost power. The transmitter resumed at 16\,209\,921, above the last counter it had used, 16\,208\,997. The receiver restored its bound, and all later frames were accepted and verified offline. A flash dump taken just before power-off showed each log as designed, with consecutive records one STRIDE (1024) apart.

\subsection{Airtime, Latency and Resources}\label{sec:resources}\label{sec:busresults}

At a 2~Mbit/s data rate, a secure frame occupies the bus for 385.5~$\mu$s on average and a baseline frame for 387.9~$\mu$s, computed from the ISO 11898-1 frame layout. Both are 64-byte frames, so the small difference comes only from stuff bits. The bare 8-byte command in an unpadded CAN~FD frame would take 153.8~$\mu$s, so the secure frame uses about 2.5 times that airtime.

An on-chip logic analyzer, added for this measurement, showed a secure frame lasting 360.2~$\mu$s at 2~Mbit/s and 192.2~$\mu$s at 5~Mbit/s. The analyzer stops at the acknowledge bit, so its figure is shorter than the computed 385.5~$\mu$s. At 5~Mbit/s the two FPGA nodes also ran the full schedule without a reject or missed slot. The replay test stayed at 2~Mbit/s because the CANable's bit timing leaves no margin for clock error at 5~Mbit/s~\cite{mutter2013icc}.

In RTL simulation, sealing and verifying a frame each take 61 cycles (Section~\ref{sec:core}), or 610~ns at 100~MHz. Compared with the plaintext baseline, the secure layer adds 128 cycles per frame on average, 63 at the sender and 65 at the receiver. That is about 1.3~$\mu$s, or 0.33\% of the frame's 385.5~$\mu$s on the bus.

Table~\ref{tab:resources} gives the post-route resources. Replay protection and persistence add 1429 LUTs and 798 flip-flops to the core, and no block RAM. The secure layer and one controller use 4286 LUTs, 20.6\% of the XC7A35T. This excludes the logic that feeds frames to the controller, which in our build sits in the 1122-LUT scenario harness.

\begin{table}[htbp]
\caption{Post-route utilization by block (XC7A35T, 100~MHz,
$\mathrm{STRIDE} = 1024$). Bold rows are subtotals; the other rows
sum to the device total, with the harness and top-level rows
counted without their child blocks.\label{tab:resources}}
\begin{tabularx}{\textwidth}{Xrrr}
\toprule
\textbf{Block} & \textbf{LUT} & \textbf{FF} & \textbf{RAMB18} \\
\midrule
Ascon-AEAD128 core (encrypt + decrypt) & 1120 & 719 & 0 \\
Freshness, transmit\,/\,receive & 57\,/\,246 & 48\,/\,48 & 0 \\
Counter persistence + SPI master & 1126 & 702 & 0 \\
\midrule
\textbf{Secure layer} & \textbf{2549} & \textbf{1517} & \textbf{0} \\
CTU CAN~FD controller, node A\,/\,B & 1737\,/\,1711 & 956\,/\,968 & 3\,/\,3 \\
\textbf{Secure layer + one controller (excl.\ harness)} & \textbf{4286} & \textbf{2473} & \textbf{3} \\
\midrule
Scenario harness & 1122 & 2676 & 0 \\
Test only: VIO + debug hub & 4071 & 7266 & 0 \\
Top-level wrapper & 36 & 105 & 0 \\
Whole device as built & 11\,226 & 13\,488 & 6 \\
\bottomrule
\end{tabularx}
\end{table}

The design meets timing at 100~MHz, with 0.404~ns to spare for setup and 0.020~ns for hold. One timing exception relaxes the paths into the controllers' bit-timing registers, which can only be written while a controller is stopped. From simulated traffic, Vivado estimates that the device draws 149~mW, 77~mW of it from switching activity.

\subsection{Comparison with Prior Work}\label{sec:discussion}

Table~\ref{tab:comparison} compares our design with the closest prior work. Our Ascon core handles both encryption and decryption in 1120 LUTs. Nguyen and Chen's core~\cite{nguyen2024ascon}, which also does hashing, needs 2464, and Alharbi et al.'s~\cite{alharbi2024ascon7series} 1756. The whole secure layer, at 2549 LUTs, is smaller than the 2988-LUT system of Nguyen and Chen, which has no replay protection. With one controller, our 4286 LUTs fall within the 3490 to 5335 LUTs of Gladis Kurian and Chen~\cite{gladiskurian2025ascon}. Their design also needs 32 block-RAM tiles and a Bloom-filter lookup of up to 1180~ns, while ours checks freshness with the tag alone.

The three FPGA designs all use Artix-7 devices, so LUT counts compare directly, although the others use the larger XC7A100T. The system totals cover different scopes. Ours includes a full controller but not the frame-feeding logic, Nguyen and Chen's includes a receive-only bus front end, and Gladis Kurian and Chen's has no bus interface. Latencies come from RTL simulation with different message lengths. Power is not compared, because ours is only a tool estimate.

\begin{table}[htbp]
\caption{Comparison with prior work. IPD = SP~800-232 initial
public draft, whose Ascon-AEAD128 parameters are identical to the
final publication. A dash means not reported or not applicable.\label{tab:comparison}}
\footnotesize
\begin{tabularx}{\textwidth}{>{\raggedright\arraybackslash}p{2.3cm}LLLL}
\toprule
 & \textbf{This work} & \textbf{N\&C'24}~\cite{nguyen2024ascon} &
\textbf{GK\&C'25}~\cite{gladiskurian2025ascon} &
\textbf{Rasheed'24}~\cite{rasheed2024tnse} \\
\midrule
Cipher & 800-232 final & v1.2 & 800-232 IPD & TinyJAMBU-128 \\
Platform & XC7A35T FPGA & XC7A100T FPGA & XC7A100T FPGA & STM32 software \\
Bus, tested on hardware & CAN FD, yes & classic CAN, yes & generic IoT, no & classic CAN, yes \\
Replay check & counter; at most 1536 frames lost per receiver reboot &
none & Bloom filter, 0.166\% false positives & context-based, detection 0.92 to 0.99 \\
Replay state kept across reboots & bounds in flash & no & no & no claim \\
Nonce sent & 2 counter bytes & not described & yes & --- \\
LUTs, core\,/\,system & 1120\,/\,4286 (secure layer + one controller) & 2464 (incl.\ hash)\,/\,2988 & 3490 to 5335 (system) & --- \\
BRAM (replay) & 0 & 0 & 32 tiles & --- \\
Encrypt\,/\allowbreak\,decrypt at 100~MHz & 610\,/\,610~ns (46-B plaintext) & 1521\,/\,1481~ns (8-B plaintext) & 510\,/\,520~ns (32-B plaintext) & --- \\
\bottomrule
\end{tabularx}
\end{table}

\section{Security Analysis and Limitations}\label{sec:security}

\subsection{Security Properties}\label{sec:keylimits}

As long as nonces never repeat and the key stays secret, the payload stays confidential and any tampered frame is rejected, within the per-key limits below. Identifiers, lengths, timing and the two counter bytes are sent in clear. A rejected frame does not change $L$. Replays are rejected only if the receiver keeps its state and $L$ stays below the guard band of Section~\ref{sec:freshness}. The receiver checks order, not age, so an old frame that was held back and never delivered could still be accepted. Applications with deadlines need their own timing check.

Ascon degrades more gracefully under nonce reuse than AES-GCM, where repeated nonces let an adversary forge tags~\cite{joux2006gcm}. If two messages share a nonce and an attacker knows one of them, the attacker can read the other only up to and including the first block where the two differ. An attack that recovers the internal state under nonce reuse is known for the older Ascon-128~\cite{wiemer2023cansec}, but none is known for Ascon-AEAD128. Ascon-AEAD128 has also been proven to keep its authenticity under nonce reuse, in the random-permutation model~\cite{mennink2023keyblinding}. We do not rely on any of this, because the counter bounds in flash prevent nonce reuse.

SP~800-232 limits how much one key may be used. A key may see at most $2^{96}$ failed decryptions and $2^{54}$ bytes of data. Each frame uses the key twice, once to encrypt and once to decrypt, so the data limit is reached after about $2^{47}$ frames. Even at $10^{5}$ frames per second, more than one CAN~FD bus can carry, that takes about 38 years. The prototype does not count key usage. A deployment must count it and change the key before either limit is reached.

\subsection{Limitations}\label{sec:limits}\label{sec:pending}

Anyone who can rewrite the external flash can restore an old bound and make counters repeat. The threat model treats this as compromise of the endpoint. Keys are installed and changed only through service access. The key-epoch byte leaves room for changing keys in the field, but key provisioning and assignment of device prefixes are out of scope.

Each transmitter reset skips between 512 and 1536 counter values, about one STRIDE on average. If the transmitter resets about 64 times without the receiver accepting a frame in between, its counter moves past the receiver's window of 65\,536 (Section~\ref{sec:freshness}). The receiver then rejects every later frame, and the prototype has no way to recover. We did not test this case.

A rejected replay still takes space in the receive buffer. With the controller's default 32-word buffer, two secure runs overran when a replay arrived just before a genuine frame, so the evaluated build uses 128 words. A production receiver must size its buffers for attack traffic.

On the physical bus, only unchanged replay was tested. Tampered frames and replays after a reboot were tested only in simulation, and delayed delivery was not tested. Both endpoints run on one FPGA with a shared clock, flash and power supply, so separate ECUs that reboot independently were not tested. We did not compare against a software Ascon implementation on a microcontroller, so we make no claim that hardware is required.

\section{Conclusions and Future Work}\label{sec:conclusion}

Our design fits a 46-byte payload, two counter bytes and a full 128-bit Ascon-AEAD128 tag into one 64-byte CAN~FD frame. The receiver rebuilds the full counter from the two bytes, and the tag confirms it. Counter bounds stored in flash keep counters from repeating across reboots. A receiver reboot costs at most 1536 frames, and the flash is written about once every 1024 frames.

On an Artix-7 XC7A35T at 100~MHz, the secure layer uses 2549 LUTs and adds about 1.3~$\mu$s per frame. On a physical bus, the plaintext baseline acted on every replayed frame, while the secure layer rejected all of them and accepted every genuine command. Over an hour of repeated reboots and a power cycle, no counter was reused.

The next step is to close the gaps of Section~\ref{sec:limits}: testing on separate ECUs with their own storage, and testing tampering, delayed delivery and replay after a reboot on the bus. The design protects one channel. On a bus with several senders and receivers, each protected identifier needs a counter at its sender and at every receiver, and because they share the key, any receiver could forge frames for that identifier. A deployment also needs key provisioning, rekeying before the per-key limits, a check that stops the node before the counter reaches the guard band, and a secure way to recover when the endpoints fall out of sync.

\section*{Author Contributions}
Conceptualization, methodology, software, validation, formal analysis, investigation, resources, data curation, writing (original draft), writing (review and editing), and visualization, K.K. and V.N.; project administration, V.N. Both authors have read and approved the manuscript.

\section*{Data and Code Availability}
The hardware sources, test benches, scenario software, implementation reports, bench captures and audit outputs supporting the reported results are available from the corresponding author on reasonable request. The CTU CAN~FD controller is available from its maintainers under its own license~\cite{ctucanfd}.

\section*{Acknowledgments}
Generative-AI tools were used in this work: Anthropic Claude Fable~5, Claude Fable~5.1, Claude Opus~4.8, Claude Opus~5 and Claude Opus~5.5, and OpenAI GPT Astra (September to November 2026), as interactive coding and writing assistants for hardware-description and test code, verification and analysis scripts, the scenario software, figures, reference checking and manuscript drafting. All designs, tests and measurements were specified, run, checked and interpreted by the authors. Reported results come from recorded simulation, synthesis and laboratory runs, not model output. The authors reviewed and edited all AI-assisted text and code and take full responsibility for the content. The authors thank the maintainers of the publicly available CTU CAN~FD controller.

\section*{Competing Interests}
K.K. is employed by optoML, Bengaluru, India, a fabless semiconductor company developing AI inference hardware. This work was carried out independently of that employment. optoML did not fund the work, provide equipment or data, or take part in the study design, analysis or manuscript. Its products are unrelated to CAN security. V.N. declares no competing interests.

\section*{Funding}
This research received no external funding. The authors bought the evaluation hardware themselves.

\renewcommand{\bibfont}{\footnotesize}\setlength{\bibsep}{0pt}
\bibliographystyle{unsrtnat}
\bibliography{references}

\end{document}